\documentclass[9pt,twocolumn,twoside]{pnas-new}

\templatetype{pnasresearcharticle} 

\newcommand{\figref}[2]{\hyperref[#1]{\ref{#1}\textit{\uppercase{#2}}}}

\title{Phase-induced topological superconductivity in a planar heterostructure}

\author[a]{Omri Lesser}
\author[b]{Andrew Saydjari}
\author[c]{Marie Wesson}
\author[b,c,1]{Amir Yacoby}
\author[a,1]{Yuval Oreg}

\affil[a]{Department of Condensed Matter Physics, Weizmann Institute of Science, Rehovot, Israel 7610001}
\affil[b]{Department of Physics, Harvard University, Cambridge, Massachusetts 02138}
\affil[c]{John A. Paulson School of Engineering and Applied Sciences, Harvard University, Cambridge, Massachusetts 02138, USA}

\leadauthor{Lesser} 

\significancestatement{The practical realization of Majorana zero modes in quasi-one-dimensional topological superconductors is greatly hindered by the need to apply strong magnetic fields. This study proposes a novel way to engineer these exotic states using only superconducting phase bias, which requires negligible magnetic fields or currents. The proposed device is experimentally accessible and robust, as we show by comprehensive theoretical modeling. Furthermore, it has the potential of providing substantially cleaner experimental signatures of Majorana zero modes than the currently available platforms, paving the way to building a topological qubit.}

\authorcontributions{All authors conceived the model. O.L. performed the simulations. A.S. and M.W. provided considerations on the experimental feasibility and performed preliminary simulations which informed the work. All authors analyzed the results and wrote the manuscript.}
\authordeclaration{The authors declare no competing interests.}
\correspondingauthor{\textsuperscript{1}To whom correspondence should be addressed. E-mails: yuval.oreg@weizmann.ac.il, yacoby@g.harvard.edu}

\keywords{topological phases of matter $|$ majorana zero modes $|$ topological superconductivity} 

\begin{abstract}
Topological superconductivity in quasi-one-dimensional systems is a novel phase of matter with possible implications for quantum computation. Despite years of effort, a definitive signature of this phase in experiments is still debated. A major cause of this ambiguity is the side effects of applying a magnetic field: induced in-gap states, vortices, and alignment issues. Here we propose a planar semiconductor-superconductor heterostructure as a platform for realizing topological superconductivity without applying a magnetic field to the two-dimensional electron gas hosting the topological state. Time-reversal symmetry is broken only by phase-biasing the proximitizing superconductors, which can be achieved using extremely small fluxes or bias currents far from the quasi-one-dimensional channel. Our platform is based on interference between this phase biasing and the phase arising from strong spin-orbit coupling in closed electron trajectories. The principle is demonstrated analytically using a simple model, and then shown numerically for realistic devices. We show a robust topological phase diagram, as well as explicit wavefunctions of Majorana zero modes. We discuss experimental issues regarding the practical implementation of our proposal, establishing it as an accessible scheme with contemporary experimental techniques. 
\end{abstract}

\begin{document}

\maketitle
\thispagestyle{firststyle}
\ifthenelse{\boolean{shortarticle}}{\ifthenelse{\boolean{singlecolumn}}{\abscontentformatted}{\abscontent}}{}

The quest for discovering novel phases of matter has seen rapid development in the past few decades, with the advent of topological quantum matter~\cite{hasan_colloquium_2010, moore_birth_2010,qi_topological_2011,bernevig_topological_2013,sato_topological_2017}.
Phases that would be equivalent under Landau's order-parameter paradigm were found to be distinguished by topological properties. 
The earliest and perhaps most salient phase characterized by its topology is the quantum Hall effect~\cite{girvin_quantum_1999}, where the chiral edge modes signal the non-trivial bulk topology~\cite{thouless_quantized_1982,stern_anyons_2008}.
It was later realized that a similar phenomenon can take place with a zero net magnetic flux~\cite{haldane_model_1988}, which opened the door to the field of Chern insulators.

Shortly after these important discoveries, the connection to superconductivity (SC) was made, by the concept of topological superconductivity. 
This novel phase of matter supports Majorana zero modes (MZMs), which are predicted to have non-Abelian exchange statistics~\cite{alicea_new_2012,nayak_non-abelian_2008}. 
It has been vigorously studied in recent years, following early ideas in one~\cite{kitaev_unpaired_2001} and two~\cite{fu_superconducting_2008} spatial dimensions. 
Further theoretical studies advanced the field from toy models to practical implementations~\cite{lutchyn_majorana_2018}, fuelled by the potential application of the exotic MZMs in quantum computation~\cite{nayak_non-abelian_2008,lahtinen_short_2017,karzig_universal_2016}.

A prominent Majorana platform is hybrid superconductor-semiconductor nanowires~\cite{lutchyn_majorana_2010,oreg_helical_2010}, where a Zeeman field combined with strong spin-orbit coupling (SOC) drive a proximitized nanowire into a topologically non-trivial phase. 
Following the theoretical proposals, several tunneling experiments~\cite{mourik_signatures_2012, das_zero-bias_2012} observed zero-bias conductance peaks as possible signatures of MZMs.
The early results triggered many subsequent studies, including on current-biased~\cite{romito_manipulating_2012-1} and disordered~\cite{brouwer_probability_2011,brouwer_topological_2011} nanowires and carbon nanotubes~\cite{sau_topological_2013,marganska_majorana_2018,lesser_topological_2020} as the one-dimensional platform.
Other important Majorana realization platforms include quantum wells subjected to an in-plane magnetic field~\cite{alicea_majorana_2010}, topological insulator-superconductor heterostructures~\cite{fu_superconducting_2008}, semiconductor-ferromagnet heterostructures~\cite{sau_generic_2010,vaitiekenas_zero-bias_2020}, chains of magnetic adatoms on spin-orbit-coupled superconductors~\cite{pientka_topological_2013,nadj-perge_observation_2014}, iron-based superconductors~\cite{xu_topological_2016,wang_evidence_2018}, and most recently full-shell proximitized nanowires~\cite{stanescu_robust_2018,vaitiekenas_flux-induced_2020}.

A different route, first taken in two important theoretical works~\cite{hell_two-dimensional_2017,pientka_topological_2017}, is based on planar semiconductor-superconductor heterostructures.
These studies proposed utilizing a phase-biased Josephson junction, forming a quasi-1D geometry, with an applied in-plane magnetic field.
Such a planar device is a highly attractive platform for the experimental realization of topological superconductivity, as it is less delicate than nanowires, does not require strict magnetic field alignment, and is easier to fabricate.
The theoretical proposals triggered two recent experiments that indeed reported zero-bias conductance peaks which may indicate the existence of MZMs~\cite{ren_topological_2019,fornieri_evidence_2019}.
However, a drawback of this platform is the need to apply an appreciable magnetic field to the device.
The applied magnetic field diminishes superconductivity and could give rise to magnetic-impurity states, hindering the detection of MZMs.

In this manuscript, we propose a route to realizing topological superconductivity in a planar geometry without applying a magnetic field in the Josephson junction. Previous proposals achieved this by using supercurrents to break time-reversal symmetry; however, the large currents required and fringing fields in the junction complicate implementation~\cite{melo_supercurrent-induced_2019}. Our proposal relies solely on phase-biasing the proximitizing superconductor, a possibility that was recently pointed out~\cite{lesser_three-phase_2021}. 
The approach is based on engineering the geometry such that the topological Aharonov-Casher phase~\cite{aharonov_topological_1984}, induced by the spin-orbit coupling, constructively interferes with the SC phase winding~\cite{fu_superconducting_2008,van_heck_single_2014,lesser_three-phase_2021}, giving rise to a zero-energy state. We would like to emphasize that the role of the phase biases in our proposal and in Fu and Kane's proposal~\cite{fu_superconducting_2008} is different. In Fu and Kane's proposal, the topological phase results from the interplay between the surface of a 3D topological insulator and a superconductor, even in the absence phase bias. The role of the phase bias is to form a discrete vortex supporting an MZM. Here the phases themselves drive the system into a topological phase in a conventional material platform with strong spin-orbit coupling. Our geometry also benefits from eliminating unbound trajectories as in the case of zig-zag junctions~\cite{laeven_enhanced_2020}. Not only does this proposal eliminate the complications of magnetic fields in the junction coexisting with the MZMs, but it is also accessible with current materials and fabrication techniques.

\section*{Minimal model and ``sweet spot"}\label{sec:minimal_model}
We shall now demonstrate the possibility of realizing MZMs in a magnetic-field-free planar geometry.
To this end, we develop a toy model that supports a ``sweet spot" with perfectly localized MZMs~\cite{fulga_adaptive_2013}. 
This is analogous to tuning the Kitaev chain to zero chemical potential and equal hopping and pairing amplitudes, which makes the edge modes perfectly localized in a single site~\cite{kitaev_unpaired_2001}.

\begin{figure*}
    \centering
    \includegraphics[width=\linewidth]{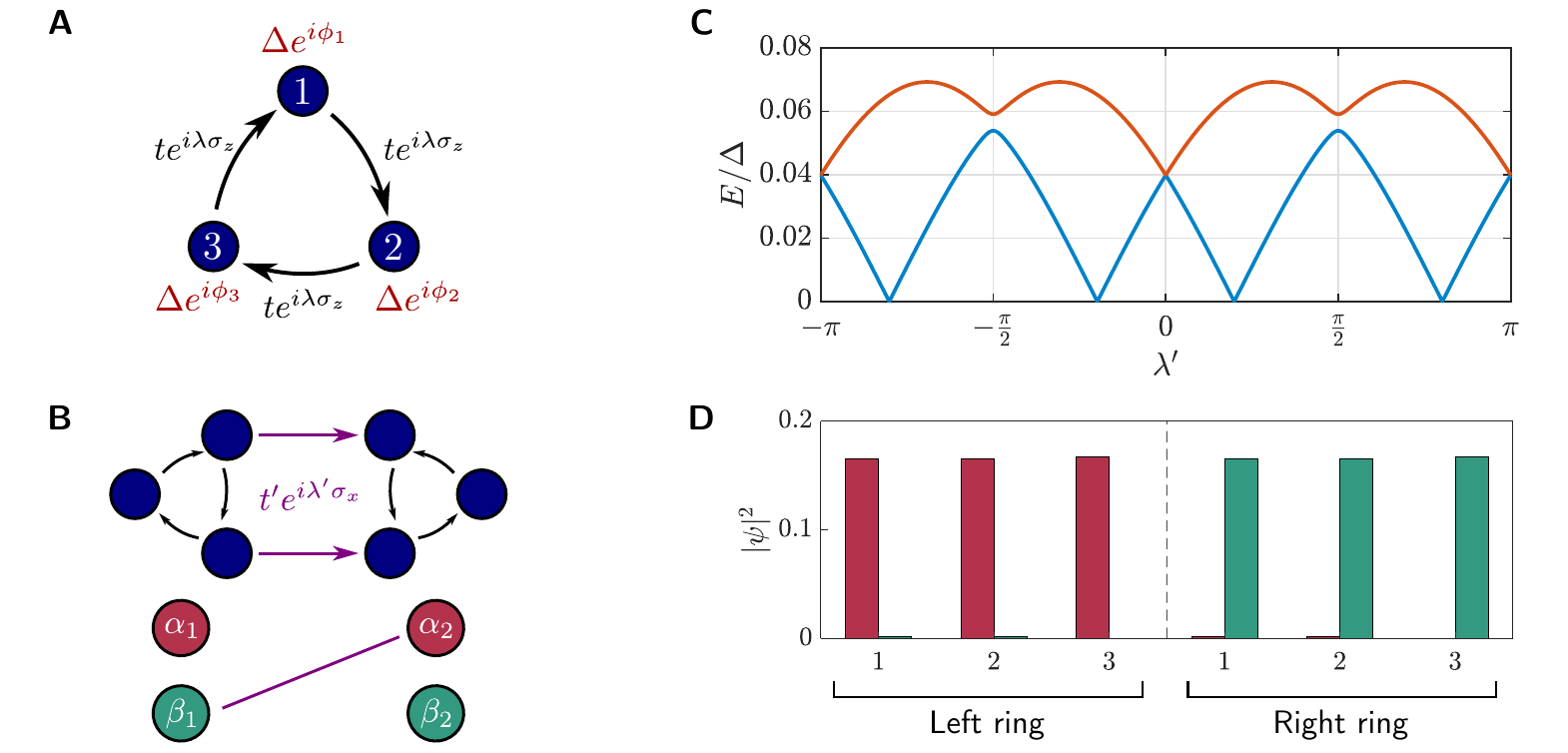}
    \caption{Toy model for realizing a ``sweet spot" relying on SC phase bias only. 
    \textit{(A})~Illustration of the ring described by \eqref{eq:H_ring_realspace} for $N=3$ sites, each connected to its nearest neighbors by a hopping amplitude $t$ through a spin-orbit-coupled medium characterized by the angle $\lambda$. Each site is proximitized by a superconductor with pair potential $\Delta$ and a different phase $\phi_n$. The parameters controlling the ring's behavior are tuned such that each ring contains two overlapping Majorana zero modes, $\alpha$ and $\beta$.
    \textit{(B)}~Two rings are coupled at two points (top). If only $\beta_1$ and $\alpha_2$ are coupled, those states are gapped out, leaving the zero-energy states $\alpha_1$, $\beta_2$ intact and localized at the left and right ring, respectively (bottom). 
    \textit{(C)}~Energies of the lowest-lying (blue) and second-lowest-lying (orange) states in the system composed of two coupled rings. For certain values of $\lambda^{\prime}$ a zero-energy state appears, with a finite gap to the second-lowest state. At $\lambda^{\prime}=0,\pm\pi$ we observe spin degeneracy.
    \textit{(D)}~Zero-energy Majorana wavefunctions at one of the special values of $\lambda^{\prime}$, each state localized in a different ring.}
    \label{fig:ring}
\end{figure*}

The basic building block of our model is a ring composed of $N\geq3$ sites, see Fig.~\figref{fig:ring}{a}.
Electrons can hop from each site to its nearest neighbors through a spin-orbit-coupled medium, and each site $n$ is proximitized by a SC with a different phase $\phi_n$.
The ring is described by the tight-binding Bogoliubov-de-Gennes (BdG) Hamiltonian 
\begin{equation}\label{eq:H_ring_realspace}
\begin{aligned}
    H&=\sum_{n=1}^{N}\left(te^{i\lambda\sigma_z}c_{n}^{\dagger}c_{n+1}-\mu c_{n}^{\dagger}c_{n}\right)\\
    &+\sum_{n=1}^{N}\left(\Delta e^{i\phi_{n}}c_{n,\uparrow}^{\dagger}c_{n,\downarrow}^{\dagger}+\text{H.c.}\right),
\end{aligned}
\end{equation}
where $\sigma$ labels the spin degree of freedom, $t$ is the hopping amplitude, $\lambda$ is the SOC angle, $\mu$ is the chemical potential, and $\Delta$ is the SC pairing potential. 
The gauge transformation $c_{n}\rightarrow c_{n}e^{i\phi_{n}/2}$ eliminates the phases from the SC terms, in exchange for adding the complex amplitude $e^{i\left(\phi_{n+1}-\phi_{n}\right)/2}$ to the hopping term from site $n+1$ to site $n$.

It is known that SC phase winding is necessary in order to obtain a zero-energy state in such a model~\cite{van_heck_single_2014}. 
For simplicity, we assume that $\phi_{n}=n\varphi$ where $\varphi=2\pi m/N$ with $m\in\mathbb{Z}$, i.e., an integer number of vortices.
With this choice $\phi_{n+1}-\phi_{n}=\varphi$ is constant, so the model becomes translationally invariant and can therefore be analyzed in momentum space.
The Bloch Hamiltonian reads 
\begin{equation}\label{eq:H_ring_kspace}
\begin{aligned}
    H&=\sum_{k,\sigma} \left[2t\cos\left(k+\frac{\varphi}{2}+\lambda\sigma\right) - \mu \right] c_{k\sigma}^{\dagger}c_{k\sigma}\\
    &+ \sum_{k}\left(\Delta c_{k\uparrow}^{\dagger}c_{-k\downarrow}^{\dagger}+\text{H.c.}\right),
\end{aligned}
\end{equation}
where $k=\frac{2\pi}{N}q,\,q=0,1,\ldots,N-1$.
As we show in Appendix~\ref{appendix:sweet_spot}, it is possible to tune the model's parameters such that two zero-energy states appear: a pair of MZMs, delocalized around the ring.
The key ingredient of the model is the interplay between the gauge-invariant topological Aharonov-Casher phase, $3 \lambda$ (arising due to electron and hole trajectories circulating the ring), and the SC phase winding.

Having established the possibility of realizing MZMs in a ring, the next step is coupling two such rings, as shown in Fig.~\figref{fig:ring}{b}. 
Each ring hosts two MZMs, and our goal is to couple the rings such that one MZM in each ring remains uncoupled, and the other two MZMs are gapped out, as illustrated at the bottom of Fig.~\figref{fig:ring}{b}. 
Since the bare MZMs are delocalized throughout their rings, the only way to achieve this goal is by interference of trajectories~\cite{creutz_aspects_2001}, and therefore the two rings must be coupled via more than one link.
In the coupling form suggested in Fig.~\figref{fig:ring}{b}, the rings are connected at two points via spin-orbit-coupled links of amplitude $t^{\prime}e^{i\lambda^{\prime}\sigma_x}$~\footnote{The $\sigma_x$ Pauli matrix appears in the inter-ring SOC because of the choice made in the original Hamiltonian \eqref{eq:H_ring_realspace} of taking all intra-ring SOC terms proportional to $\sigma_z$, which is a local rotation of the spin.}.

By choosing the SC phases at the two rings and controlling the inter-ring SOC angle $\lambda^{\prime}$, it is possible to tune into a ``sweet spot" where one MZM in each ring remains intact. 
We show this explicitly in Appendix~\ref{appendix:sweet_spot} by projecting the coupling Hamiltonian to the low-energy subspace and making sure that one MZM in each ring is left intact.
This situation is demonstrated in Fig.~\figref{fig:ring}{c}, where we plot the energies of the lowest and second-lowest states as a function of $\lambda^{\prime}$.
At special values of $\lambda^{\prime}$, the lowest energy is zero and the second-lowest energy is finite, implying the existence of two MZMs, each localized in a different ring, with a gap to excitations. 
If we now concatenate these double-ring building blocks on the plane, the localization length of the Majorana edge modes will be one ring independent of the overall length of the chain. 
We have therefore achieved a topological superconducting phase by tuning the three superconducting phases, without the Zeeman effect. A sweet spot is obtained by setting $\mu$, $\Delta$, $\lambda$, $\lambda^{\prime}$ to proper values.

\section*{Realistic model}\label{sec:realistic_model}

How can one realize a phase similar to the ``sweet spot" in a realistic experiment? 
The key point is using a geometry that supports closed orbits with a non-zero Aharonov-Casher phase~\cite{aharonov_topological_1984} and SC phase winding~\cite{lesser_three-phase_2021}.
Here we demonstrate this idea using a planar semiconductor-superconductor heterostructure and analyze the topological phase diagram as a function of easily controlled parameters---the SC phases and the chemical potential.

\begin{figure}
    \centering
    \includegraphics[width=0.8\linewidth]{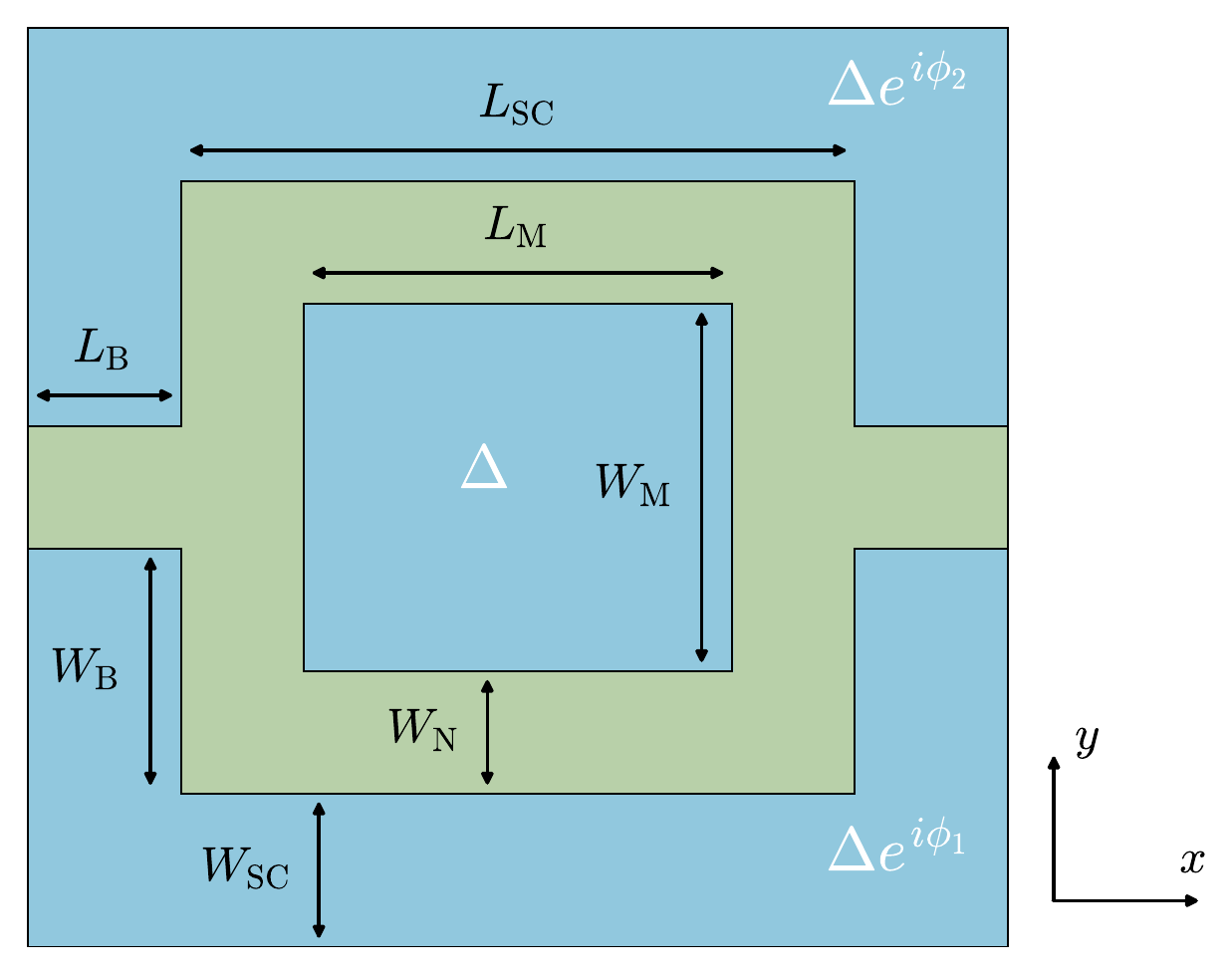}
    \caption{Unit cell of the proposed geometry for realizing topological superconductivity in a planar heterostructure using phase bias only. Blue regions are covered by SCs, whereas the green region is an uncovered 2DEG. The SC phases are $0$ at the middle SC, $\phi_1$ and the bottom SC, and $\phi_2$ at the top SC. The geometric parameters are indicated on the schematic. When $W_{\rm B}>W_{\rm N}$ straight trajectories that do not encounter SCs at all are eliminated, thus facilitating a large topological gap~\cite{laeven_enhanced_2020}. The unit cell is repeated along the $x$ direction.}
    \label{fig:geometry}
\end{figure}

We consider the unit-cell geometry depicted in Fig.~\ref{fig:geometry}. 
The system is made of a two-dimensional electron gas (2DEG) with Rashba SOC, partially covered by SCs with fixed phases. 
The main difference between this configuration and the original proposals that rely on a Zeeman field~\cite{hell_two-dimensional_2017,pientka_topological_2017} is the addition of a third SC with phase-control. 
The non-straight features, characterized by the lengths $\left(W_{\rm B},L_{\rm B}\right)$, will help stabilize the topological phase by increasing the energy gap (along the lines of Ref.~\cite{laeven_enhanced_2020}).
In what follows we will show that when the SC phases wind, the system can be driven into a topological SC phase akin to the one realized in the toy model, even at zero applied Zeeman field.

The heterostructure is described by the Hamiltonian
\begin{equation}\label{eq:H_2DEG_SC}
\begin{aligned}
    H=&\left[\frac{\hbar^{2}}{2m^{*}}\left(k_{x}^{2}+k_{y}^{2}\right)+\hbar\alpha\left(\sigma_{y}k_{x}-\sigma_{x}k_{y}\right)-\mu\right]\tau_{z}\\
    &+{\rm Re}\Delta(x,y)\tau_{x}+{\rm Im}\Delta^*(x,y)\tau_{y},
\end{aligned}
\end{equation}
where $k_x,k_y$ are the momenta along the $x,y$ directions, $m^{*}$ is the effective electron mass, $\alpha$ is the Rashba SOC parameter, $\mu$ is the chemical potential, $\Delta$ is the local superconducting pairing potential, and $\tau$, $\sigma$ Pauli matrices act in particle-hole and spin spaces, respectively.
The magnitude of $\Delta(x,y)$ is taken to be a constant $\Delta$ in the proximitized regions, and zero in the non-proximitized regions.
The SC phases are 0 at the middle SC, $\phi_1$ at the bottom SC, and $\phi_2$ at the top SC, as indicated in Fig.~\ref{fig:geometry}.
The corresponding Nambu spinor is 
$\vec{\Psi}=\left( \psi_{\uparrow},\psi_{\downarrow},\psi_{\downarrow}^{\dagger},-\psi_{\uparrow}^{\dagger} \right)^{\rm T}$, 
where $\psi_{s}$ annihilates an electron of spin $s$ along the $z$ axis.

The system is assumed be finite along the $y$ direction, and the unit cell is repeated along the $x$ direction.
For the numerical simulations, the Hamiltonian~\eqref{eq:H_2DEG_SC} is discretized on a square lattice of (unless specified otherwise, the lattice spacing is $a=10\,{\rm nm}$).
Further details on the tight-binding model are given in Appendix~\ref{appendix:tight_binding}.
We note that while the geometry proposed in Fig.~\ref{fig:geometry} is the focus of this study, it is merely an example of the general concept. As long as the geometry supports trajectories with a discrete vortex and a spin-orbit phase, it is expected to work as well. We have numerically demonstrated a topological phase transition in several similar geometries.

\section*{Topological phase diagram}\label{sec:phase_diagram}

In this section we present the full topological phase diagram of the proposed system.
The Hamiltonian~\eqref{eq:H_2DEG_SC} belongs to symmetry class D~~\cite{altland_nonstandard_1997,schnyder_classification_2008,kitaev_periodic_2009}, with the particle-hole operator being $\mathcal{P}=\tau_{y}\sigma_{y}\mathcal{K}$ ($\mathcal{K}$ is complex conjugation).
Enhancement to the BDI class, as in the case of a single planar Josephson junction~\cite{pientka_signatures_2013}, is generically prevented by the existence of two distinct phase differences, and only occurs at fine-tuned conditions.
In (quasi) one dimension, systems in symmetry class D possess a $\mathbb{Z}_2$ topological invariant that we label $\mathcal{Q}$, which takes the values $+1,-1$ for the trivial and topological phases, respectively.
We calculate the invariant by attaching two leads to a finite-size system and invoking the scattering formula $\mathcal{Q}={\rm det}r$, where $r$ is the reflection matrix from one of the leads to itself~\cite{fulga_scattering_2012}. The scattering calculations are performed using the Kwant software package~\cite{groth_kwant_2014}.

Apart from the topological invariant, the key property of the system is the energy gap. 
To find the gap, we impose periodic boundary conditions along $x$, such that the momentum $k_x$ is a good quantum number.
We then densely scan the Brillouin zone and calculate the lowest energy at each value of $k_x$.

\begin{figure}
    \centering
    \includegraphics[width=\linewidth]{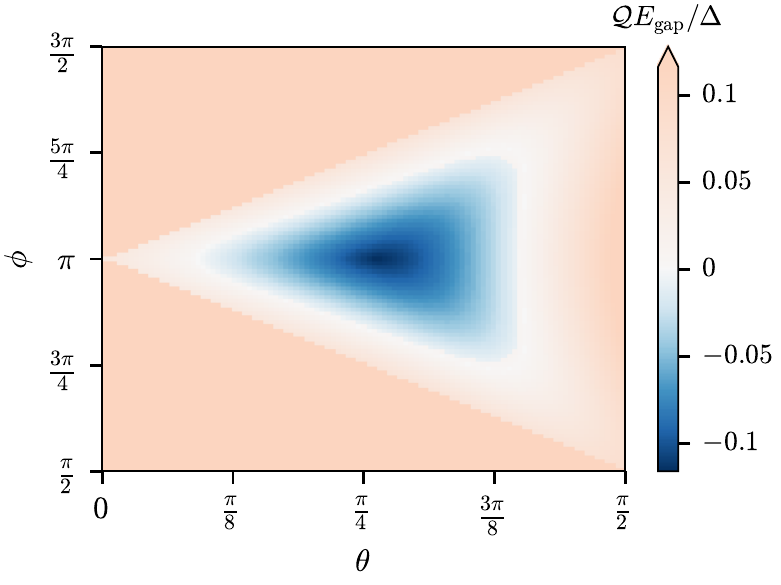}
    \caption{Topological phase diagram of the model \eqref{eq:H_2DEG_SC}, showing the topological invariant $\mathcal{Q}$ which is $-1$ ($+1$) in the topological (trivial) phase, multiplied by the energy gap. The phase diagram is shown as a function of $\theta=\left(\phi_1-\phi_2\right)/2$ and $\phi=\left(\phi_1+\phi_2\right)/2$. The physical parameters are $\mu=0.26\,{\rm meV}$, $\alpha=23\,{\rm meVnm}$, and the geometric parameters are $L_{\rm SC}=320\,{\rm nm}$, $W_{\rm SC}=50\,{\rm nm}$, $L_{\rm B}=50\,{\rm nm}$, $W_{\rm B}=80\,{\rm nm}$, $L_{\rm M}=240\,{\rm nm}$, $W_{\rm M}=120\,{\rm nm}$, $W_{\rm N}=40\,{\rm nm}$. See~\cite{interactive_figure} for an interactive version of this figure.}
    \label{fig:phase_diagram}
\end{figure}

Fig.~\ref{fig:phase_diagram} shows the resulting phase diagram as a function of the phase differences $\theta=\left(\phi_1-\phi_2\right)/2$ and $\phi=\left(\phi_1+\phi_2\right)/2$.
The system only supports a topological phase when the SC phases wind, which is evident from the phase diagram: $Q=-1$ only in certain regions inside the triangle $\pi-\theta<\phi<\pi+\theta$ (for $0\leq\theta\leq\pi/2$).
This triangle sets the optimal phase boundaries we can expect in such a system~\cite{van_heck_single_2014,lesser_three-phase_2021}, and the topological region we obtain here is not far from that.
Furthermore, we get a substantial topological gap of over $10\%$ of the induced SC gap $\Delta$.
Such a large energy gap has important implications on making the MZM localization length short, and on enabling reliable experimental detection.
The maximal value of the gap is obtained near the center of the topological region.

In this structure, closed trajectories in which the electron scatters off all three SCs seem vital for the existence of the topological phase~\cite{lesser_three-phase_2021}.
They are present also when the discontinuous rectangles of the middle SC are connected by a thin SC, thus becoming continuous.
Then, normal tunneling or crossed Andreev tunneling through the thin segment will enable such trajectories.
We have verified numerically that this is indeed possible, but in our checks this came at price of lowering the topological gap.
Notice though that in a completely straight geometry, i.e., $W_{\rm B}=0$, it is impossible to obtain a topological phase---the required closed orbits do not exist.

\begin{figure}
    \centering
    \includegraphics[width=\linewidth]{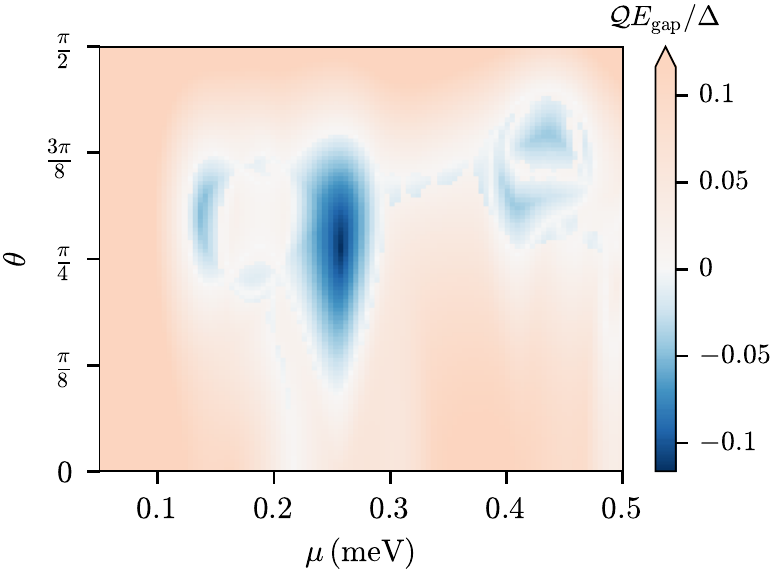}
    \caption{Topological phase diagram of the model \eqref{eq:H_2DEG_SC}, as a function of the chemical potential $\mu$ and the phase difference $\theta$, for $\phi=\pi$. The parameters are the same as in Fig.~\ref{fig:phase_diagram}.}
    \label{fig:mu_theta_diagram}
\end{figure}

To demonstrate the robustness of the topological phase, we examine its stability to variations in the chemical potential and phase bias.
In Fig.~\ref{fig:mu_theta_diagram} we show the topological phase diagram as a function of $\mu$ and $\theta$, fixing $\phi=\pi$.
We find that the topological phase persists for an appreciable range of parameters, indicating that no delicate fine-tuning is necessary for the system to support MZMs.
Stability to such variations is of paramount practical importance, as elaborated in the experimental considerations section.

The geometric parameters are chosen here according to the rule of thumb $L\Delta\approx\alpha$~\cite{lesser_three-phase_2021}, where $L$ is a typical length in the uncovered 2DEG region. Given different 2DEG and SC material, one should tune the lengths to approximately match this condition (see Appendix~\ref{appendix:niobium} and the interactive figure~\cite{interactive_figure}).

\section*{Real-space analysis and Majorana wavefunctions}\label{sec:real_space}
In the previous section, we have established the existence of a topological phase by calculating the $\mathbb{Z}_2$ invariant. 
We now turn our focus to the real-space hallmark of the topological phase, which is the existence of localized zero-energy Majorana end states.
These states are the truly tangible manifestation of the topological nature of the phase, and experimentally they reveal themselves as zero-bias conductance peaks in tunneling measurements~\cite{alicea_majorana_2010,mourik_signatures_2012,das_zero-bias_2012,ren_topological_2019,fornieri_evidence_2019}.

\begin{figure}
    \centering
    \includegraphics[width=\linewidth]{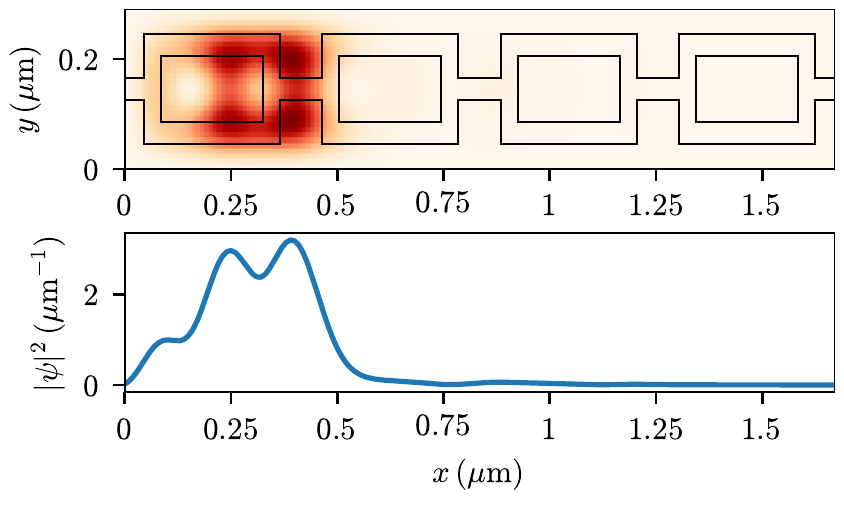}
    \caption{Real-space simulation of Hamiltonian~\eqref{eq:H_2DEG_SC} with geometry proposed in Fig.~\ref{fig:geometry}, in the topological phase.
    Top: wavefunction of the zero-energy Majorana state localized at the left edge. The wavefunction is almost completely localized at the first unit cell. The black borders indicate the boundary between the SC and the 2DEG. Notice that the MZM wavefunction has significant support beneath the SCs; we find that such behavior occurs for a wide range of parameters but not always.  Bottom: the above wavefunction integrated over $y$ shows exponential decay along $x$. An exponential fit  yields a localization length of $\xi_{\rm M}\approx 0.5\,\mu{\rm m}$.
    The simulation included 10 unit cells, and for clarity only the 4 left-most ones are shown. The other MZM is localized at the right edge of the device.
    The parameters used are the same as in Fig.~\ref{fig:phase_diagram}, with $\phi=\pi$, $\theta=\pi/4$.}
    \label{fig:real_space}
\end{figure}

A typical MZM wavefunction is shown in Fig.~\ref{fig:real_space}, both as a 2D heat map and as a 1D curve (integrated along the transverse direction $y$). 
The parameters are chosen well within the topological phase, as inferred from Fig.~\ref{fig:phase_diagram}. 
As we expect, two states of near zero energy appear in this regime, localized at opposite edges of the device.
The MZM wavefunctions are concentrated a bit more on the normal 2DEG regions than on the SC regions, though its noticeable leakage into the SC region indicates a strong proximity effect.

\begin{figure}
    \centering
    \includegraphics[width=\linewidth]{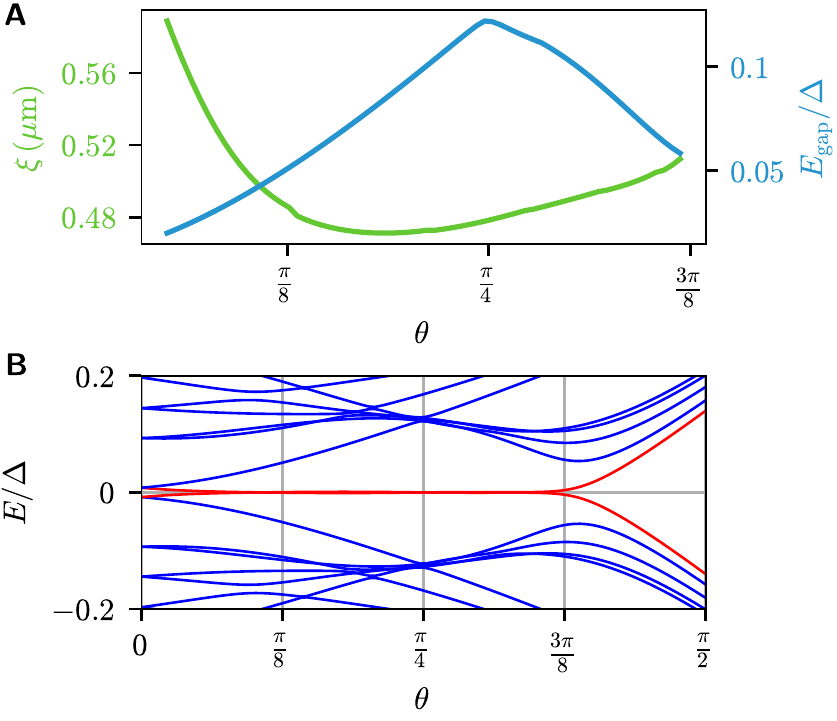}
    \caption{\textit{(A)}~Localization length (green) and energy gap (blue) in the topological phase.
    The localization length is roughly inversely related to the energy gap, as expected by the continuum approximation $\xi\approx\hbar v_{\rm F}/E_{\rm gap}$, where $v_{\rm F}$ is the Fermi velocity.
    \textit{(B)}~Lowest-lying BdG energies as a function of the phase difference $\theta$ in a finite system.
    The topological phase transition is signaled by a closing and re-opening of the energy gap, while leaving two Majorana states (red) at zero energy.
    The parameters used are the same as in Fig.~\ref{fig:phase_diagram}, with $\phi=\pi$ and 6 unit cells.}
    \label{fig:rs_loc_and_spectrum}
\end{figure}

Remarkably, the Majorana states are localized over one unit cell, which is about $0.5\, \mu{\rm m}$ long. A smaller unit cell for the same spin-orbit energy would deviate from the optimal geometry, leading to a smaller topological gap.
This localization length also agrees with the continuum approximation $\xi\approx\hbar v_{\rm F}/E_{\rm gap}$, where $v_{\rm F}$ is the Fermi velocity.
The relation between the localization length and the energy gap is further demonstrated in Fig.~\figref{fig:rs_loc_and_spectrum}{a}, by calculating both of them as a function of $\theta$ inside the topological phase.

We close this section with another explicit demonstration of the topological phase transition.
Fig.~\figref{fig:rs_loc_and_spectrum}{b} shows the evolution of the lowest-lying energies as a function of $\theta$ in a finite system of 6 units cells~\footnote{Notice that $\theta$ is chosen for convenience, and a similar plot as a function of $\mu$ or $\phi$ may also be obtained.}.
The energy gap closes and re-opens, leaving a pair of MZMs bound to zero energy.
This signals the transition from the trivial to the topological phase.
The topological gap is maximized near $\theta=\pi/4$, and then closes and re-opens again near $\theta=3\pi/8$, this time leaving no in-gap states, indicating that the system returns to the topologically trivial phase.
Notice the agreement between this real-space calculation and the $\phi=\pi$ cut of the phase diagram shown in Fig.~\ref{fig:phase_diagram}.

\section*{Experimental considerations}\label{sec:experimental}
This proposal for phase-only topological superconductivity can apply to a general class of material platforms: large spin-orbit semiconducting quantum wells such as HgTe, InAs, and InSb which have well developed fabrication procedures~\cite{bendias2018high,lee2019transport,ke2019ballistic}. While induced-superconductivity has been demonstrated in all three platforms, we focus on HgTe below due to its high mobility ($\mu$ = 100--800 $\times$10$^{3}$ cm$^{2}$V$^{-1}$s$^{-1}$) and lack of surface states.

\begin{figure}
    \centering
    \includegraphics[width=\linewidth]{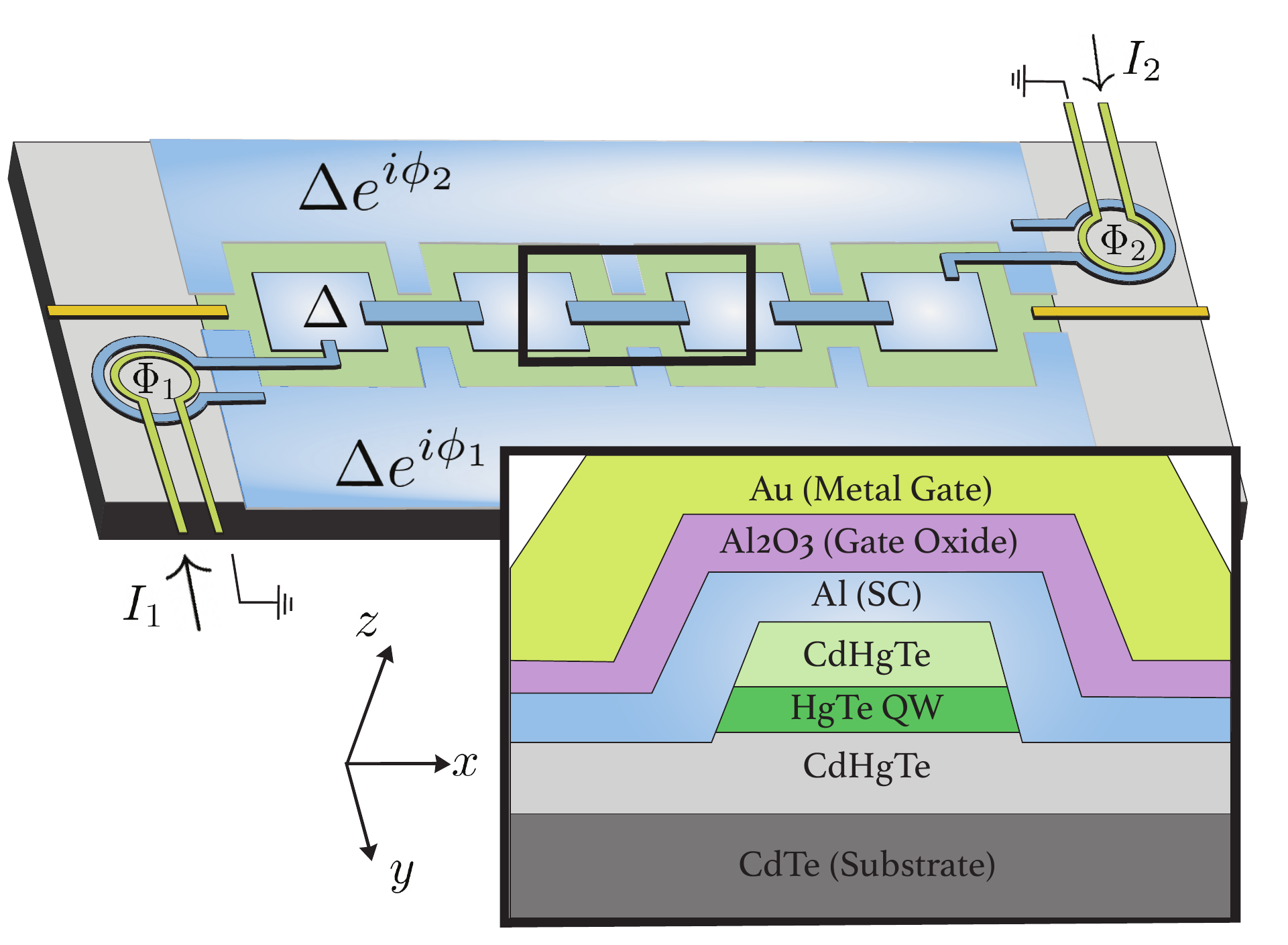}
    \caption{Feasible device fabrication layout. Light blue regions represent superconductors which are deposited after etching to make contact to the quantum well (QW). Superconducting connectors between center SC islands are deposited in the same step using an airbridge technique or a separate step milling through native oxides which form after the first SC deposition. The two phase degrees of freedom are set independently by current loops which eliminates even fringing fields from the junction region which hosts the MZMs. 2DEG is removed outside the junction region. The whole device is covered in a metal on oxide electrostatic gate for tuning the density/chemical potential (not shown in top view).  Conductance tunnel probes, shown in orange, are included for detecting MZMs at the ends of the device.
    }
    \label{fig:experimentalHgTe}
\end{figure}

The minimum feature sizes of the proposed geometry (Fig.~\ref{fig:geometry}) can be achieved by standard electron beam lithography ($\geq$ 30 nm). See caption of Fig.~\ref{fig:experimentalHgTe} for brief outline of the proposed fabrication procedure. Practical geometry constraints are also placed by the material. The width $W$ between superconducting contacts must be smaller than the coherence length of the induced superconductivity in the Josephson junction which is limited to $\zeta_T \sim \hbar v_{\rm F}$/$2\pi k_{\rm B} T$. While the induced coherence length provides a loose bound on the junction width to $\leq$ 50 $\mu$m at 20 mK, this bound quickly becomes strict at higher temperatures. Further, in order to work in the quasi-ballistic regime, the mean free path $l_e$ must be at least on the order of $W$. For HgTe, we have $l_e\sim 1.6\,\mu\rm{m}$ for $\mu\sim 100\times10^{3}\,{\rm cm}^2 {\rm V}^{-1} {\rm s}^{-1}$ and $n\sim 1\times10^{12}\,{\rm  cm}^{-2}$. Since there is no competing drive for larger junction widths, as in the case of the in-plane field proposal \cite{pientka_topological_2017,hell_two-dimensional_2017}, this upper bound need not be saturated thus suggesting our proposal will be more robust to lower-quality semiconducting materials.

With these upper and lower bounds from the material in mind, we still have to determine the optimal operational dimensions of the device. Dimensional analysis suggests that a typical coherence length set by $\alpha/\Delta$ should be of the order of the typical lengths of the device $L$. Ref.~\cite{lesser_three-phase_2021} analyzed a simplified tight-binding model of a related (non-planar) system, and showed that this rule of thumb indeed leads to an optimal situation, with a maximal topological region in the $\theta$--$\phi$ plane of Fig.~\ref{fig:phase_diagram}. For fixed $\Delta$, this rule of thumb, $L \Delta \approx \alpha$, implies that for larger spin-orbit coupling, larger unit cells are required to achieve the largest topological gap. In contrast to fully 1D proposals, here the confinement of the MZM is set by the size of the unit cell, rather than just the topological gap as is expected in a continuum setting. For moderate $\alpha = 23$~meV~nm with the superconducting gap of aluminium, this condition fixes $L \sim$ 1 $\mu$m. For fixed $\alpha$, increasing the superconducting gap shortens this length scale. Advantageously, HgTe growth can be tuned from negligible toward the very large spin-orbit splitting up to 30 meV \cite{gui_giantSO_2004, rothe_SOterms_2010}, with $\alpha$ as large as 120~meV~nm, which can be matched to the superconductor used to balance fabrication limitations on dimension and optimize confinement. Notice that even for the case of Al, we saturate lithographic precision at $\alpha \sim 10$~meV~nm, where $E_{\rm SO} \sim 1.5\,{\rm meV} \geq \Delta$, suggesting extraordinary confinement is possible. 

The most difficult aspect of implementing the phase-only proposal is tuning stably into the topological phase. Our parameter space consists of geometrical lengths, 2DEG density, spin-orbit strength, and superconducting phases. Geometrical lengths are set during the fabrication and can only be made uniform to within $10$ nm. Thus, for a fixed geometry, we require our tunable parameters to bring us to a topological regime. In addition, the existence of a topological phase cannot be sensitive to small variations in lengths from unit cell to unit cell, or even within a unit cell. We have demonstrated this insensitivity in~\cite{interactive_figure}.

While back-gates on semiconductor quantum wells are possible \cite{baenninger2012fabrication}, none of the platforms discussed above have implemented this technology while maintaining high mobility. As a result, control over the 2DEG density and spin-orbit strength are both coupled to the potential of the top gate. Gated, the mean density of HgTe can be tuned from 4$\times$10$^{11}$ cm$^{-2}$ to 1$\times$10$^{12}$ cm$^{-2}$ while the spin-orbit splitting can be tuned up to 30 meV \cite{gui_giantSO_2004,reuther_gatedefined_2013}. In Figs.~\ref{fig:mu_theta_diagram} and \ref{fig:rs_loc_and_spectrum}, the simulated density is approximately $1-6 \times 10^{9}$ and $0.6-4 \times 10^{10}$ cm$^{-2}$ respectively, with stable regions corresponding to approximately $1 \times 10^{9}$ and $1.4 \times 10^{10}$ cm$^{-2}$ are found. Since these densities are much less than typical semiconductor densities and the stable regions are on the order of the fluctuations of the gate potential, we simulate more realistic conditions in Fig.~\ref{sfig:high_density}, with a density of $5 \times 10^{11}$cm$^{-2}$ showing stable regions of width $5 \times 10^{10}$ cm$^{-2}$ which should be achievable with current experimental platforms. 

The superconducting phases are easily tuned during operation of this device. The size of the phase-biasing loops and the current limits set the phase resolution and stability. A 10--20 $\mu {\rm m}^2$ loop can be fabricated with ease, is robust to sub-Gauss offsets in the ambient magnetic field, and can be biased with currents $\leq$ 150 $\mu {\rm A}$ without causing significant Joule heating. With these parameters, a current resolution of 10 nA corresponds to a flux precision of $2 \times 10^{-4} \phi_0$. As shown in Figs.~\ref{fig:mu_theta_diagram}, \ref{sfig:niobium}, \ref{sfig:high_density}, this precision is more than sufficient to tune stably into the topological regime.

As shown in Fig.~\ref{fig:experimentalHgTe}, conductance tunnel probes can be used to detect MZMs at the ends of the junction.  In many MZM tunneling experiments, the topological gap is close to the energy resolution of the tunnel probe ($10$ $\mu$eV). The large gaps in our proposal, 100 $\mu$eV for Nb, should mitigate this issue. The type-II nature of large-gap superconductors often complicates measurements due to small lower critical magnetic fields $H_{c1}$ which allow flux penetration, motion, and trapping during device operation in external magnetic fields. However, the lack of critical magnetic fields should be an ideal use case for large-gap superconductors \footnote{Type-II superconductors also have the disadvantage for phase-based proposals of short-coherence lengths. Here the relevant coherence length should be that in the semiconductor, $\zeta_T \sim \hbar v_{\rm F}$/$2\pi k_{\rm B} T$, and thus avoid this problem.}. The fact that we do not have an external field decreasing the parent gap of the superconductor should further improve the MZM visibility during tunneling experiments to the end of the junction. 

\section*{Conclusion}\label{sec:conclusion}

We have shown that MZMs can arise in a phase-controlled planar semiconductor-superconductor device without applying a Zeeman field.
At the heart of our scheme lies interference between Aharonov-Casher phase~\cite{aharonov_topological_1984}, stemming from the spin-orbit interaction, and SC phase winding.
As shown exactly by means of the toy model and then numerically with a realistic model, proper tuning of these two phases drives the system into a topological superconducting state.

Several important advantages of our proposal are worth mentioning.
It enjoys all the benefits of planar geometry compared to nanowires, in particular easier fabrication and the ability to generalize to a Majorana network.
Moreover, since the proposal is interference-based rather than Zeeman-based, various limitations related to magnetic fields are lifted.
In the original planar Josephson junction proposals~\cite{pientka_topological_2017,hell_two-dimensional_2017}, the Zeeman field required to make the system topological is of the order of the Thouless energy, which scales like the inverse of the junction's width.
Therefore making the junction narrower -- which is desirable to achieve more ballistic transport -- comes at the price of applying a stronger magnetic field.
Here the magnetic field is not needed, making the above consideration irrelevant and facilitating the use of extremely narrow junctions.
Furthermore, MZM-related experiments typically favor the use of Al due to it being a type-I SC that does not trap flux.
Our proposal enables the use of larger-gap type-II SCs, such as Nb, since flux trapping is no longer a problem, thus opening the door to robust large-gap MZMs.

The discussion in the previous section shows that our proposal is realistic under existing experimental technologies.
The entire class of large spin-orbit semiconductors are viable platforms, and importantly no topological materials are required.
As we have shown, the MZM can be localized to within a single unit cell, whose size is set by the device geometry.
The smallest physical dimension is set by fabrication limitations and the SOC strength (by the $L\Delta\approx\alpha$ criterion).
Thus, in the limit of large SOC, the maximal topological gap can coincide with an extremely short localization length.

In the future, it will be interesting to explore possible extensions of the design principles we suggest and investigate here in quasi one-dimensional systems to two dimensions, e.g., via a coupled-wire construction. This is expected to lead to the formation of a two-dimensional chiral superconducting state.

\acknow{We are grateful to A. Akhmerov, A. Melo, and B. A. Bernevig for illuminating discussions.
The work at Weizmann was supported by the European Union's Horizon 2020 research and innovation programme (Grant Agreement LEGOTOP No. 788715), the DFG (CRC/Transregio 183, EI 519/7-1), ISF Quantum Science and Technology (2074/19), the BSF and NSF (2018643).
A.S. gratefully acknowledges support by a National Science Foundation Graduate Research Fellowship (DGE-1745303). M.W. is supported by the DoD NDSEG Fellowship. A.Y., A.S., and M.W. are supported by the Quantum Science Center (QSC), a National Quantum Information Science Research Center of the U.S. Department of Energy (DOE), the Gordon and Betty Moore Foundation through Grant GBMF 9468, the National Science Foundation under Grant No. DMR-1708688, and the STC Center for Integrated Quantum Materials, NSF Grant No. DMR-1231319.}

\showacknow{} 

\bibliography{pnas-library}

\newpage
\appendix
\setcounter{equation}{0}
\renewcommand{\theequation}{S\arabic{equation}}
\setcounter{figure}{0}
\renewcommand{\thefigure}{S\arabic{figure}}
\setcounter{subsection}{0}
\renewcommand{\thesubsection}{S.\Roman{subsection}}
\section*{Supporting Information}

\subsection{Full analysis of the sweet spot}\label{appendix:sweet_spot}

Here we show the complete derivation of the conditions for a zero-energy state in an $N$-sites ring, find their wavefunctions, and calculate the coupling between zero modes of adjacent rings.

We begin from the Bloch Hamiltonian describing the ring, \eqref{eq:H_ring_kspace} of the main text.
Labeling $\epsilon_{k\sigma}=-\mu+2t\cos\left(k+\frac{\varphi}{2}+\lambda\sigma\right)$, we write the BdG Hamiltonian in matrix form:
\begin{gather}
    H=\frac{1}{2}\sum_{k}\vec{\Psi}_{k}^{\dagger}\mathcal{H}\left(k\right)\vec{\Psi}_{k},\\
    \mathcal{H}\left(k\right)=\begin{pmatrix}\epsilon_{k\uparrow} & 0 & \Delta & 0\\
    0 & \epsilon_{k\downarrow} & 0 & \Delta\\
    \Delta & 0 & -\epsilon_{-k\downarrow} & 0\\
    0 & \Delta & 0 & -\epsilon_{-k\uparrow}
    \end{pmatrix},
\end{gather}
where the Nambu spinor is $\vec{\Psi}_{k}=\left(c_{k\uparrow},c_{k\downarrow},c_{-k\downarrow}^{\dagger},-c_{-k\uparrow}^{\dagger}\right)^{\rm T}$. 
The condition for a zero-energy state is 
\begin{equation}
    \det\mathcal{H}\left(k\right)=\left(\Delta^{2}+\epsilon_{k\uparrow}\epsilon_{-k\downarrow}\right)\left(\Delta^{2}+\epsilon_{k\downarrow}\epsilon_{-k\uparrow}\right)=0,
\end{equation}
for some value of $k$.
Choosing a specific pair $\left(k,\sigma\right)$, there are many choices of parameters fulfilling the requirement $\Delta^{2}+\epsilon_{k\sigma}\epsilon_{-k,-\sigma}=0$.
For convenience, let us consider the simple choice $\epsilon_{k\sigma}=-\epsilon_{-k,-\sigma}=\Delta$:
\begin{subequations}
\begin{equation}
    -\mu+2t\cos\left(k+\lambda\sigma+\frac{\varphi}{2}\right)=\Delta
\end{equation}
\begin{equation}
    -\mu+2t\cos\left(k+\lambda\sigma-\frac{\varphi}{2}\right)=-\Delta.
\end{equation}
\end{subequations}
Adding the two equations, we get an expression for the chemical potential, 
\begin{equation}
    \mu=2t\cos\left(\frac{\varphi}{2}\right)\sqrt{1-\left(\frac{\Delta}{2t\sin\left(\frac{\varphi}{2}\right)}\right)^{2}}.
\end{equation}
For simplicity let us set $\mu=0$, thus requiring $\Delta=\pm2t\sin\left(\frac{\varphi}{2}\right)$.

As in the main text we focus on $N=3$, so the two non-trivial choices for $\varphi$ are $\pm\frac{2\pi}{3}$. 
We therefore obtain $\Delta=\pm\sqrt{3}t$. 
We still need to tune the SOC parameter $\lambda$ in order to guarantee the existence of a zero-energy state. This is done by explicitly writing
\begin{equation}
    \epsilon_{k\sigma}=\Delta\ \Rightarrow\ \cos\left(k+\lambda\sigma+\frac{\varphi}{2}\right)=\frac{\Delta}{2t}=\frac{\sqrt{3}}{2}=\cos\left(\frac{\pi}{6}\right).
\end{equation}
Focusing on the case of a single vortex $\varphi=\frac{2\pi}{3}$ and choosing $k=0$, $\sigma=\uparrow$ for our zero-energy state, we find $\lambda=-\frac{\pi}{2}$ (up to integer multiples of $2\pi$). 

We have ensured the existence of a zero-energy state in the ring. To find its wavefunction, we need to solve the equation
\begin{equation}
   \begin{pmatrix}\epsilon_{k,\sigma} & \Delta\\
    \Delta & -\epsilon_{-k-\sigma}
    \end{pmatrix}\begin{pmatrix}c_{k\sigma}\\
    \sigma c_{-k-\sigma}^{\dagger}
    \end{pmatrix}=0. 
\end{equation}
But since we set $\epsilon_{k\sigma}=-\epsilon_{-k,-\sigma}=\Delta$, the equation takes the simple form 
\begin{equation}
    \Delta\begin{pmatrix}1 & 1\\
    1 & 1
    \end{pmatrix}\begin{pmatrix}c_{k\sigma}\\
    \sigma c_{-k-\sigma}^{\dagger}
    \end{pmatrix}=0.
\end{equation}
The eigen-energies are clearly $2\Delta$ (for the symmetric combination) and $0$ (for the anti-symmetric combination).
The zero-energy state's wavefunction is
\begin{equation}
    \psi_{0\uparrow}=\frac{1}{\sqrt{2}}\left(c_{0\uparrow}- c_{0\downarrow}^{\dagger}\right).
\end{equation}
Having found a zero-energy state (and its particle-hole partner), we can construct Majorana wavefunctions:
\begin{subequations}
\begin{equation}
    \alpha=\frac{1}{\sqrt{2}}\left(\psi_{0\uparrow}+\psi_{0\uparrow}^{\dagger}\right)=\frac{1}{2}\left(c_{0\uparrow}- c_{0\downarrow}^{\dagger}\right)+\text{H.c.}
\end{equation}
\begin{equation}
    \beta=\frac{i}{\sqrt{2}}\left(\psi_{0\uparrow}-\psi_{0\uparrow}^{\dagger}\right)=\frac{i}{2}\left(c_{0\uparrow}+ c_{0\downarrow}^{\dagger}\right)+\text{H.c.}
\end{equation}
\end{subequations}

We now discuss the coupling of two rings, as shown in Fig.~\figref{fig:ring}{b} of the main text. 
As we have shown, the parameters of each ring can be tuned such that it hosts a pair of MZMs at a particular $k,\sigma$ combination. 
Flipping both $k$ and $\sigma$ corresponds to flipping the vorticity, $\varphi\rightarrow-\varphi$. 
Let us then choose $k=0$ in both rings, and $\varphi=\pm\frac{2\pi}{3}$ opposite in the two rings (vortex--anti-vortex configuration), while keeping $\lambda=-\frac{\pi}{6}$ uniform. 
This leads to opposite $\sigma$ values for the MZMs at the two rings. 
In the real-space fermionic basis, we take the coupling to be 
\begin{equation}
    V=\begin{pmatrix}1 & 0 & 0\\
    0 & 1 & 0\\
    0 & 0 & 0
\end{pmatrix}
\otimes t^{\prime}e^{i\lambda^{\prime}\sigma_{x}}e^{i\frac{\pi}{3}},
\end{equation}
where the global $e^{i\frac{\pi}{3}}$ factor corresponds to a phase bias between the rings (this phase bias is not an essential ingredient and it is used here for analytical tractability; numerical simulations show that the same outcome is found by simply rotating the rings with respect to each other).

Projecting the coupling matrix to the basis of the MZM wavefunctions, we obtain the following coupling matrix between the MZMs at the left and right ring:
\begin{equation}
    \mathcal{M}=it^{\prime}\begin{pmatrix}0 & \cos\lambda^{\prime}+\sqrt{3}\sin\lambda^{\prime}\\
    \cos\lambda^{\prime}-\sqrt{3}\sin\lambda^{\prime} & 0
\end{pmatrix}.
\end{equation}
The rows of $\mathcal{M}$ correspond to $\alpha,\beta$ in ring 1, and its columns correspond to $\alpha,\beta$ in ring 2, so our choice automatically eliminates the $\alpha\alpha$, $\beta\beta$ couplings.
If we now choose $\lambda^{\prime}=\lambda=-\frac{\pi}{6}$, i.e., keep the SOC amplitude \emph{uniform} throughout the system, we obtain 
\begin{equation}
    \mathcal{M}=\begin{pmatrix}0 & 0\\
    i\sqrt{3} t^{\prime} & 0
\end{pmatrix}.
\end{equation}
This means that we have a sweet spot, as illustrated in Fig.~\figref{fig:ring}{b} of the main text.

So far we specified to the case $N=3$. 
In the $N=4$ case (or in general any even $N$), the quantization conditions support $k$ values for which $\sin k=0$ or $\cos k=0$, allowing some more flexibility in tuning the parameters. 
For example, if we repeat the calculation from before for $N=4$ with $k_{1}=-k_{2}=\frac{\pi}{2}$ and $\sigma_{1}=-\sigma_{2}=1$ (which is again a vortex--anti-vortex configuration; we need to set $\Delta=\sqrt{2}t$ to get $\mu=0$), we get the following Majorana coupling matrix:
\begin{equation}
\mathcal{M}=it^{\prime}\begin{pmatrix}0 & \cos\left(\lambda^{\prime}-\frac{\pi}{4}\right)\\
\cos\left(\lambda^{\prime}+\frac{\pi}{4}\right) & 0
\end{pmatrix}.
\end{equation}
By choosing $\lambda^{\prime}=\frac{\pi}{4}$, we can eliminate one of the off-diagonal elements and arrive at a sweet spot. 
In this case, a phase bias of $\frac{\pi}{4}$ is needed between the rings.

\subsection{Details of the tight-binding simulation}\label{appendix:tight_binding}

Consider a 2DEG with Rashba spin-orbit coupling:
\begin{equation}
    H\left(k_{x},k_{y}\right)=\frac{\hbar^{2}}{2m^{*}}\left(k_{x}^{2}+k_{y}^{2}\right)+\hbar\alpha\left(\sigma_{y}k_{x}-\sigma_{x}k_{y}\right)-\mu,
\end{equation}
where $m^{*}$ is the effective electron mass, $\alpha$ is the SOC parameter, and $\mu$ is the chemical potential.
At $k_y=0$, this Hamiltonian describes two shifted parabolas intersecting at $k_x=0$.
The shift in the momentum gives the SOC momentum and length:
\begin{equation}
    k_{\text{SO}}=\frac{m^{*}\alpha}{\hbar},\, \ell_{\text{SO}}=\frac{\hbar}{m^{*}\alpha}.
\end{equation}
The Fermi momentum for spin $\sigma$ is given by 
\begin{equation}
    k_{\text{F},\sigma}=k_{\text{F}}^{(0)}-\sigma k_{\text{SO}}=\frac{\sqrt{2m^{*}\mu}}{\hbar}-\frac{\sigma m^{*}\alpha}{\hbar},
\end{equation}
and the Fermi velocity, which is unaffected by the SOC, is $v_{\rm F}=\sqrt{2\mu/m^{*}}$.

We discretize this continuum model on a lattice of spacing $a$. The resulting nearest-neighbors tight-binding Hamiltonian is 
\begin{equation}
    \begin{aligned}
    H_{\text{TB}}\left(k_{x},k_{y}\right)= & -\mu_{\text{TB}}+2t_{\text{TB}}\left[2-\cos\left(k_{x}a\right)-\cos\left(k_{y}a\right)\right]\\
 & +2\alpha_{\text{TB}}\left[\sigma_{y}\sin\left(k_{x}a\right)-\sigma_{x}\sin\left(k_{y}a\right)\right],
    \end{aligned}
\end{equation}
where $t_{\text{TB}}$ is the nearest-neighbors hopping amplitude, and $\mu_{\rm TB}$, $\alpha_{\rm TB}$ are the tight-binding analogs of $\mu$, $\alpha$.
In order to match the continuum and tight-binding descriptions, we focus on $k_y=0$:
\begin{equation}
\begin{aligned}
    H_{\text{TB}}\left(k_{x},k_{y}=0\right) &=  -\mu_{\text{TB}}+2t_{\text{TB}}\left[1-\cos\left(k_{x}a\right)\right]\\
    &+2\alpha_{\text{TB}}\sigma_{y}\sin\left(k_{x}a\right).
\end{aligned}
\end{equation}
Our goal is to correctly describe the physics near the Fermi points, so we require that the Fermi momenta and the Fermi velocities match in the two descriptions.
Due to the symmetry of the model, it is enough to look at a single spin species $\sigma=+1$.

The conditions for matching the Fermi momenta are
\begin{align}
\begin{aligned}
    \mu =&	-\mu_{\text{TB}}+2t_{\text{TB}}\left[1-\cos\left(k_{\text{F}}^{\left(0\right)}a+k_{\text{SO}}a\right)\right]\\&+2\alpha_{\text{TB}}\sin\left(k_{\text{F}}^{\left(0\right)}a+k_{\text{SO}}a\right),
\end{aligned}
    \\
\begin{aligned}    
    \mu =&	-\mu_{\text{TB}}+2t_{\text{TB}}\left[1-\cos\left(-k_{\text{F}}^{\left(0\right)}a+k_{\text{SO}}a\right)\right]\\&+2\alpha_{\text{TB}}\sin\left(-k_{\text{F}}^{\left(0\right)}a+k_{\text{SO}}a\right).
\end{aligned}
\end{align}
The matching of the Fermi velocities is imposed by (notice the signs)
\begin{align}
\begin{aligned}
    \frac{\hbar v_{\text{F}}^{\left(0\right)}}{a}=&	2t_{\text{TB}}\sin\left(k_{\text{F}}^{\left(0\right)}a+k_{\text{SO}}a\right)\\&+2\alpha_{\text{TB}}\cos\left(k_{\text{F}}^{\left(0\right)}a+k_{\text{SO}}a\right),
\end{aligned}
\\
\begin{aligned}
    -\frac{\hbar v_{\text{F}}^{\left(0\right)}}{a}=&	2t_{\text{TB}}\sin\left(-k_{\text{F}}^{\left(0\right)}a+k_{\text{SO}}a\right)\\&+2\alpha_{\text{TB}}\cos\left(-k_{\text{F}}^{\left(0\right)}a+k_{\text{SO}}a\right).
\end{aligned}
\end{align}
Let us arrange these four equations in a matrix, 
\begin{equation}
\begin{gathered}
\underbrace{\begin{pmatrix}-1 & 2\left[1-\cos\left(k_{\text{F}}^{\left(0\right)}a+k_{\text{SO}}a\right)\right] & 2\sin\left(k_{\text{F}}^{\left(0\right)}a+k_{\text{SO}}a\right)\\
-1 & 2\left[1-\cos\left(-k_{\text{F}}^{\left(0\right)}a+k_{\text{SO}}a\right)\right] & 2\sin\left(-k_{\text{F}}^{\left(0\right)}a+k_{\text{SO}}a\right)\\
0 & 2\sin\left(k_{\text{F}}^{\left(0\right)}a+k_{\text{SO}}a\right) & 2\cos\left(k_{\text{F}}^{\left(0\right)}a+k_{\text{SO}}a\right)\\
0 & 2\sin\left(-k_{\text{F}}^{\left(0\right)}a+k_{\text{SO}}a\right) & 2\cos\left(-k_{\text{F}}^{\left(0\right)}a+k_{\text{SO}}a\right)
\end{pmatrix}}_{A}
\\
\times\begin{pmatrix}\mu_{\text{TB}}\\
t_{\text{TB}}\\
\alpha_{\text{TB}}
\end{pmatrix}=\underbrace{\begin{pmatrix}\mu\\
\mu\\
\frac{\hbar v_{\text{F}}^{\left(0\right)}}{a}\\
-\frac{\hbar v_{\text{F}}^{\left(0\right)}}{a}
\end{pmatrix}}_{\vec{b}}.
\end{gathered}
\end{equation}
This set of linear equations is \emph{overdetermined}. The best least-squares approximate solution is given by~\cite{anton_elementary_2010} 
\begin{equation}
\begin{pmatrix}\mu_{\text{TB}}\\
t_{\text{TB}}\\
\alpha_{\text{TB}}
\end{pmatrix}=\left(A^{\text{T}}A\right)^{-1}A^{\text{T}}\vec{b}.
\end{equation}
Substituting $A$ and $\vec{b}$ we obtain 
\begin{subequations}\label{seq:TB_matching}
\begin{equation}
    \mu_{\text{TB}} = -\mu+\frac{\hbar v_{\text{F}}^{\left(0\right)}}{a}\left[\frac{\cos\left(k_{\text{SO}}a\right)}{\sin\left(k_{\text{F}}^{\left(0\right)}a\right)}-\cot\left(k_{\text{F}}^{\left(0\right)}a\right)\right],
\end{equation}
\begin{equation}
    t_{\text{TB}} = \frac{\hbar v_{\text{F}}^{\left(0\right)}}{2a}\frac{\cos\left(k_{\text{SO}}a\right)}{\sin\left(k_{\text{F}}^{\left(0\right)}a\right)},
 \end{equation}   
\begin{equation}
    \alpha_{\text{TB}} = -\frac{\hbar v_{\text{F}}^{\left(0\right)}}{2a}\frac{\sin\left(k_{\text{SO}}a\right)}{\sin\left(k_{\text{F}}^{\left(0\right)}a\right)}.
\end{equation}
\end{subequations}

\begin{figure}
    \centering
    \includegraphics[width=\linewidth]{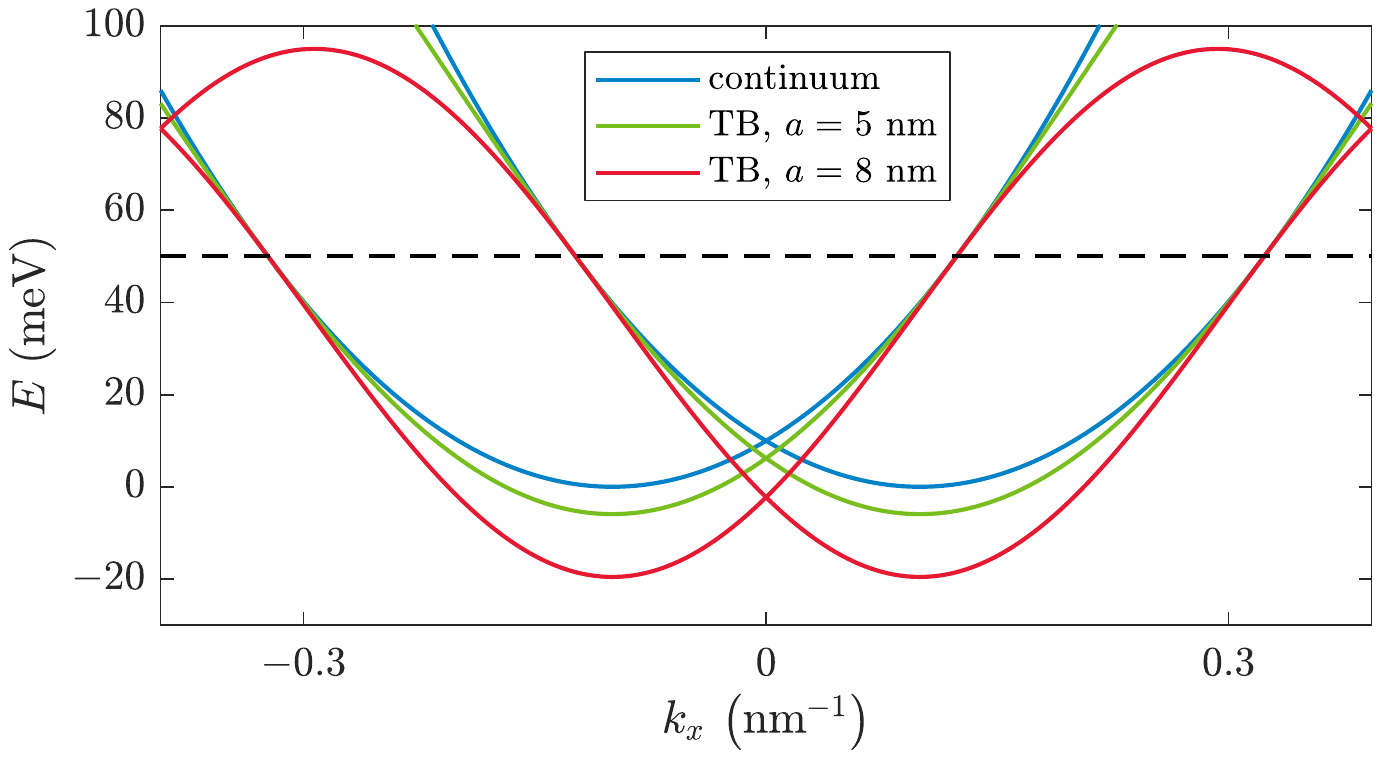}
    \caption{Continuum band structure of a HgTe 2DEG at $k_y=0$ (blue) matched by tight-binding models with lattice spacings $a=5\,{\rm nm}$ (green) and $a=8\,{\rm nm}$ (red), according to \eqref{seq:TB_matching}. The parameters are $m^{*}=0.038m_e$, $\hbar\alpha=200\,{\rm meV\,nm}$, and $\mu=50\, {\rm meV}$, corresponding to the density $n=7.9\times10^{11}\text{ cm}^{-2}$. Both models correctly capture the physics near the four Fermi points, though the $a=5\,{\rm nm}$ model does so for a larger range of energies.}
    \label{sfig:band_structure}
\end{figure}

Using this prescription, it is possible to match a tight-binding description to a continuum model, with a large lattice spacing.
The lattice spacing is limited by the requirement 
$\left(k_{\text{F}}^{\left(0\right)}+k_{\text{SO}}\right)a<\pi$, such that all Fermi points lie in the first Brillouin zone.
In Fig.~\ref{sfig:band_structure} we show such a matching for $a=5\,{\rm nm}$ and $a=8\,{\rm nm}$, where the chemical potential is such that $a$ must be smaller than $9.7\,{\rm nm}$. 
It is evident that even the model with $a=8\,{\rm nm}$, which is quite close to the upper bound on $a$, captures the continuum spectrum within about $5\,{\rm meV}$ from the Fermi energy, which is much larger than the scale set by the SC pairing potential and the typical temperature in relevant experiments.

\subsection{Results for Niobium}\label{appendix:niobium}

In this appendix, we show that Nb can be used as the proximitizing superconductor.
The SC gap of Nb is $\Delta\approx1\,{\rm meV}$ -- much larger than that of Al.
As we show here, the topological gap in our device when using Nb can be a significant fraction of this gap, making the MZMs even more robust and localized.

\begin{figure*}
    \centering
    \includegraphics[width=\linewidth]{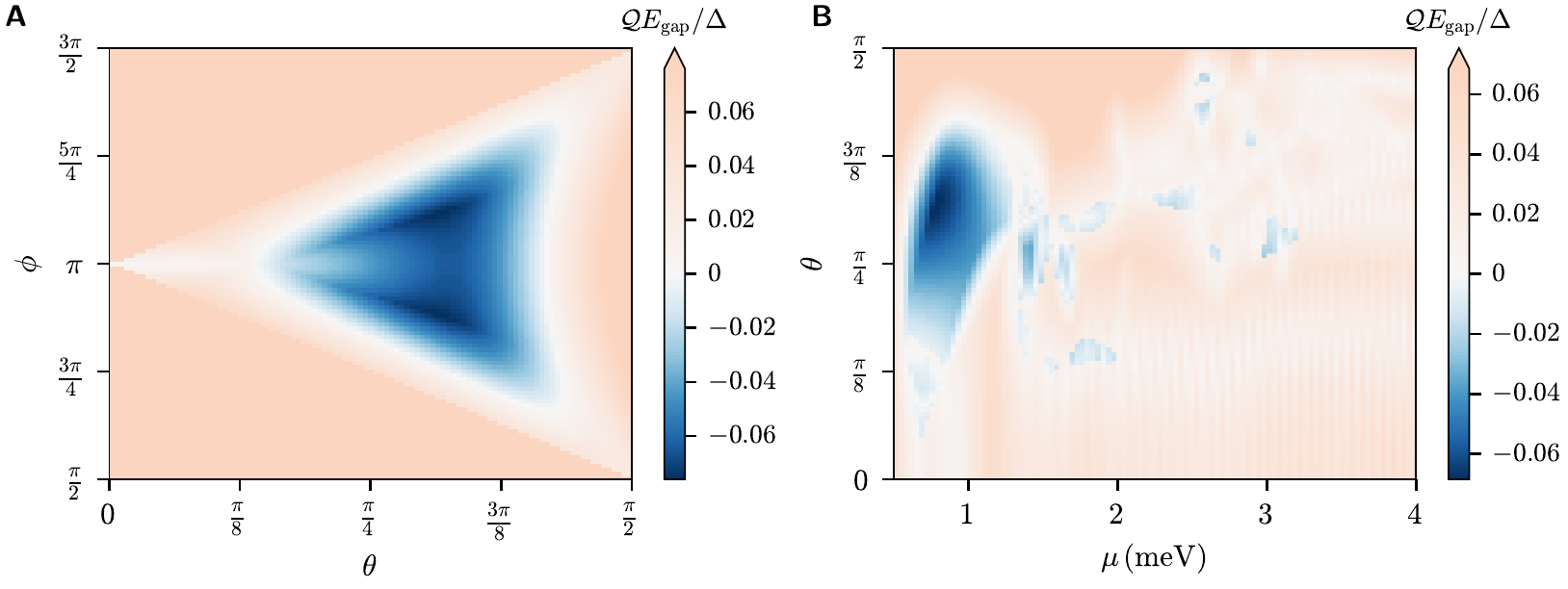}
    \caption{Nb ($\Delta=1{\rm meV}$) as the proximitizing SC. \textit{(A)}~Phase diagram as a function of the SC phase differences $\theta$ and $\phi$ for $\mu=0.87\,{\rm meV}$, showing a substantial topological region and a large topological gap. \textit{(B)}~Phase diagram as a function of the chemical potential $\mu$ and the phase difference $\theta$ for $\phi=\pi$. The SOC parameter is $\alpha=29\,{\rm meVnm}$, and the geometric parameters are $L_{\rm SC}=108\,{\rm nm}$, $W_{\rm SC}=30\,{\rm nm}$, $L_{\rm B}=30\,{\rm nm}$, $W_{\rm B}=48\,{\rm nm}$, $L_{\rm M}=60\,{\rm nm}$, $W_{\rm M}=72\,{\rm nm}$, $W_{\rm N}=24\,{\rm nm}$. The lattice constant is $a=6\,{\rm nm}$.}
    \label{sfig:niobium}
\end{figure*}

In Fig.~\figref{sfig:niobium}{a} we show the phase diagram as a function of $\theta$ and $\phi$, the analog of Fig.~\ref{fig:phase_diagram} which was calculated for Al.
The topological region of the phase diagram is quite large, and the maximal topological gap is $~0.08\Delta$ -- quite similar to the fraction obtained for Al.
This shows that when the SOC is strong enough such that the spin-orbit energy does not limit the topological gap, we can take advantage of the large proximity-induced gap.
Fig.~\figref{sfig:niobium}{b} demonstrate the chemical potential dependence (it is the analog of Fig.~\ref{fig:mu_theta_diagram} which was calculated for Al). 
This diagram shows stability over an appreciable range of electron densities, which is experimentally important. 
We note that the physical dimensions of the device should be smaller compared to Al, in line with the general rule $L\Delta\approx\alpha$ (since $\Delta$ is larger).

\begin{figure}
    \centering
    \includegraphics[width=\linewidth]{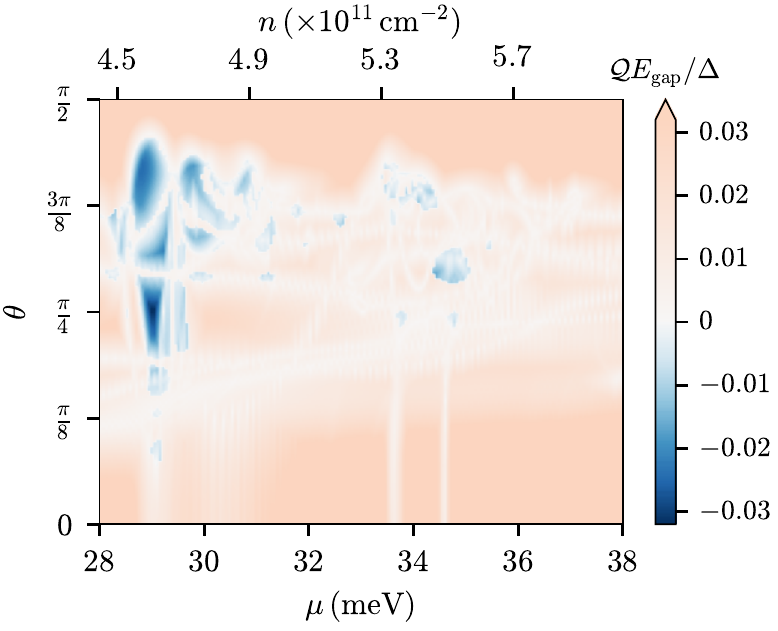}
    \caption{High-density phase diagram as a function of the chemical potential and the phase difference $\theta$ for $\phi=\pi$, as in Figs.~\ref{fig:mu_theta_diagram} and \figref{sfig:niobium}{b}. The diagram exhibits topological regions over significant ranges of density. To emphasize the experimental accessibility, we also indicate the densities on the figure. The SOC parameter is $\alpha=103\,{\rm meVnm}$, and the geometric parameters are $L_{\rm SC}=320\,{\rm nm}$, $W_{\rm SC}=48\,{\rm nm}$, $L_{\rm B}=48\,{\rm nm}$, $W_{\rm B}=80\,{\rm nm}$, $L_{\rm M}=240\,{\rm nm}$, $W_{\rm M}=120\,{\rm nm}$, $W_{\rm N}=40\,{\rm nm}$. The lattice constant is $a=8\,{\rm nm}$.}
    \label{sfig:high_density}
\end{figure}

Furthermore, to test our proposal under realistic semiconductor densities, we performed simulations at higher $\mu$ values with Nb. As detailed in Appendix~\ref{appendix:tight_binding}, this requires increasing the number of lattice points. The resulting phase diagram is shown in Fig.~\ref{sfig:high_density}. We find stable topological regions even at these realistic densities, implying that our proposal is relevant for existing semiconductor platforms. Notice that the maximum topological gap we obtain is $0.03\Delta$ -- a bit smaller than the gaps of Fig.~\ref{sfig:niobium} (though it can be improved to about $0.045\Delta$ by deviating from $\phi=\pi$). That is because the larger lattice makes the numerical optimization more computationally challenging, and therefore we only performed rudimentary optimization to check the validity of our platform. 

\end{document}